# Strong Performance Enhancement in Lead-Halide Perovskite Solar Cells through Rapid, Atmospheric Deposition of *n*-type Buffer Layer Oxides

*Ravi D. Raninga,[†] Robert A. Jagt,[†] Solène Béchu, Tahmida N. Huq, Mark Nikolka, Yen-Hung Lin, Mengyao Sun, Zewei Li, Wen Li, Muriel Bouttemy, Mathieu Frégnaux, Henry J. Snaith, Philip Schulz, Judith L. MacManus-Driscoll, Robert L. Z. Hoye\**

R. D. Raninga, R. A. Jagt, T. N. Huq, Prof. J. L. MacManus-Driscoll, Dr. R. L. Z. Hoye
Department of Materials Science and Metallurgy, University of Cambridge, 27 Charles Babbage Road, Cambridge CB3 0FS, UK

Dr. S. Béchu, Dr. M. Bouttemy, Dr. M. Frégnaux, Dr. P. Schulz
IPVF, Institut Photovoltaïque d'Île de France, 18 boulevard Thomas Gobert, 91120, Palaiseau, France

Dr. S. Béchu, Dr. M. Bouttemy, Dr. M. Frégnaux
Institut Lavoisier de Versailles (ILV), Université de Versailles Saint-Quentin en Yvelines, Université Paris-Saclay CNRS, 45 avenue des Etats-Unis, 78035 Versailles, France

Dr. M. Nikolka, Z. Li
Cavendish Laboratory, University of Cambridge, JJ Thomson Ave, Cambridge CB3 0HE, UK

Dr. Y.-H. Lin, Dr. W. Li, Prof. H. J. Snaith
Clarendon Laboratory, University of Oxford, Parks Road, Oxford OX1 3PU, UK

M. Sun
Department of Materials Science, Fudan University, Shanghai 200433, People's Republic of China

Dr. W. Li
Institute of Flexible Electronics, Northwestern Polytechnical University, Xi'an 710072, People's Republic of China

Dr. P. Schulz
CNRS, Institut Photovoltaïque d'Île de France (IPVF), UMR 9006, 18 boulevard Thomas Gobert, 91120, Palaiseau, France

Dr. R. L. Z. Hoye
Department of Materials, Imperial College London, Exhibition Road, London SW7 2AZ, UK
Email: r.hoye@imperial.ac.uk

[†] These authors contributed equally

Thin (approximately 10 nm) oxide buffer layers grown over lead-halide perovskite device stacks are critical for protecting the perovskite against mechanical and environmental damage.

However, the limited perovskite stability restricts the processing methods and temperatures (≤110 °C) that can be used to deposit the oxide overlayers, with the latter limiting the electronic properties of the oxides achievable. In this work, we demonstrate an alternative to existing methods that can grow pinhole-free $TiO_x$ ($x$ = 2.00±0.05) films with the requisite thickness in <1 min without vacuum. This technique is atmospheric pressure chemical vapor deposition (AP-CVD). The rapid but soft deposition enables growth temperatures of ≥180 °C to be used to coat the perovskite. This is ≥70 °C higher than achievable by current methods and results in more conductive $TiO_x$ films, boosting solar cell efficiencies by >2%. Likewise, when AP-CVD $SnO_x$ ($x ≈ 2$) is grown on perovskites, there is also minimal damage to the perovskite beneath. The $SnO_x$ layer is pinhole-free and conformal, which reduces shunting in devices, and increases steady-state efficiencies from 16.5% (no $SnO_x$) to 19.4% (60 nm $SnO_x$), with fill factors reaching 84%. This work shows AP-CVD to be a versatile technique for growing oxides on thermally-sensitive materials.

## 1. Introduction

Lead-halide perovskites are attracting substantial attention for energy and optoelectronic applications, owing to rapidly rising efficiencies in devices, including photovoltaics,[1] light-emitting diodes,[2,3] X-ray detectors[4] and photodetectors.[5] These materials have the stoichiometry $ABX_3$, in which A is a monovalent cation (*e.g.*, methylammonium or $MA^+$), B a divalent cation (*i.e.*, $Pb^{2+}$) and X a halide (*e.g.*, $I^-$). As such, lead-halide perovskites are versatile materials with band gaps that can be tuned over the entire visible light range by changing the composition.[6] These materials exhibit an exceptional tolerance to point defects, *i.e.*, the lead-halide perovskites can achieve long charge-carrier lifetimes and high luminescence quantum yields, despite being grown from solution with high densities of vacancies and interstitials.[7,8] An important restriction, however, is the limited thermal, environmental and mechanical stability of lead-halide perovskites. $MAPbI_3$, in particular, is thermodynamically unstable[9] and thin films have been reported to degrade in ambient air at room temperature within days.[10] Improved thermal stability has been achieved in 'triple-cation' perovskites, in which the $MA^+$ is replaced with a mixture of $Cs^+$, formamidinium ($FA^+$) and $MA^+$.[11] However, the photovoltaic performance of triple-cation perovskites still decreases from 20% to 18% after 250 h continuous operation at room temperature in an inert nitrogen environment.[11] Similarly, the low mechanical stability of the perovskites results in films being damaged when bombarded by energetic particles, such as during sputter-deposition.[12] Recently, the use of compact oxides grown over the perovskite and organic charge transport layer has led to significant improvements in stability. By depositing a 50 nm thick layer of Al-doped ZnO nanoparticles on *p-i-n* structured perovskite solar cells, Bush *et al.* successfully protected the perovskite from mechanical damage when growing a transparent indium tin oxide (ITO) electrode on top by sputter-deposition. The resultant semi-transparent perovskite top-cells were used in four-terminal tandems with silicon.[12] However, the nanoparticle layer needed to be thick to be pinhole-free, leading to series resistance losses.[12,13] This limitation was subsequently overcome by using atomic layer deposition (ALD) to grow a pinhole-free, thin bilayer of 4 nm $SnO_x$ and 2 nm zinc tin oxide.[13] ALD has also recently been used to grow a 18–40 nm thick layer of amorphous $TiO_2$ over triple-cation perovskite, which protected the perovskite for 2 h in acidic electrolytes, allowing them to be used as photocathodes for water splitting.[14] Whilst ALD is now used in the industrial-scale production of silicon solar cells, deposition times are long due to the low growth rate and need to carry out the reaction under vacuum. For example, growing 7 nm ALD $SnO_2$ at 100 °C takes 1 h.[15] Sputter-deposition grows oxides faster but usually damages the perovskite.[12,13,16] A recent work showed that perovskites with sputtered ZnO buffer layers were less efficient than similar devices with spin-coated ZnO nanoparticle buffer layers (12.5% *vs*. 16.1% respectively).[17]

Although a wide range of deposition methods and processing conditions can be used to grow oxide charge transport layers beneath the perovskite, only a small selection of these methods and a narrow range of growth temperatures have been shown to be compatible with growing oxides over the perovskite. Apart from spin-coating colloidal nanocrystals, ALD and sputter-deposition, thermal evaporation has also been used, but mainly for high work function *n*-type oxides to extract holes rather than low work function oxides for extracting electrons. In order to manufacture oxide buffer layers on a large scale for device applications (*e.g.*, top-cells or photocathodes), the growth technique should: 1) yield conductive oxides without damaging the perovskite, 2) result in pinhole-free and dense films, 3) feature rapid but uniform film growth, 4) be scalable, and 5) produce films that are conformal to textured substrates. The latter is particularly important for applications in tandem solar cells with silicon, in which the highest efficiencies have been achieved with front-textured silicon, requiring conformal perovskite, charge transport and oxide buffer layers.[18,19] However, the current techniques used for growing oxide buffer layers are not able to achieve all requirements simultaneously (**Table 1**). Whilst dense, conformal and uniform oxide films can be achieved by ALD and thermal evaporation, both techniques are slow, vacuum-based methods with limited throughput. By contrast, spin-coating oxide nanoparticles is a fast and facile method on a lab-scale without requiring vacuum, but is not conformal and is challenging to scale-up to the module-level. A particular challenge with all growth methods is that the thermal instability of the lead-halide perovskite limits the maximum processing temperature which can be used for the deposition of oxides over perovskite films. For example, Palmstrom *et al.* found that cesium-formamidinium-based perovskites (more thermally-stable than $MAPbI_3$) degraded when ALD $SnO_2$ was deposited at 110 °C or above.[15] As shown in Table S1, SI, oxides grown by any of the current techniques have typically been processed at up to 100 °C, and in many cases at room temperature. But this limits the electronic properties of the oxides achievable, since higher mobilities and, presumably in some cases, improved band alignment of the oxide with the perovskite is achieved at higher deposition or annealing temperatures.[20–22] Variants of ALD have also been proposed that can grow pinhole-free and conformal oxides more rapidly. These are pulsed chemical vapor deposition (pulsed-CVD, in which the purge times in ALD are reduced)[15] and spatial atomic layer deposition (in which the metal precursor and oxidant are spatially-separated).[23,24] However, Palmstrom *et al.* found that cesium-formamidinium-based perovskites still degrade when depositing $SnO_2$ buffer layers at 110 °C or above by pulsed-CVD.[15] Riedl *et al.* found that when increasing the temperature for growing $SnO_2$ on perovskites by spatial ALD to above 80 °C, there was still a decrease in the efficiency of $MAPbI_3$ solar cells.[25] These methods are therefore not suited to overcome the challenges of the current techniques for growing oxide buffer layers on perovskites.

**Table 1.** Summary of the key requirements for growing oxide buffer layers on lead-halide perovskite device stacks and how much each oxide growth method fulfils each requirement[a]

| Deposition method/Requirement | Wide processing window[b] | Pinhole-free | Fast growth | Scalable | Conformal |
|---|---|---|---|---|---|
| Spin-coating oxide nanoparticles | ✗ | – | ✓ | ✗ | ✗ |
| Atomic layer deposition | ✗ | ✓ | – | ✓ | ✓ |
| Sputter deposition | ✗ | ✓ | ✓ | ✓ | ✓ |
| Thermal evaporation | ✗ | ✓ | – | – | ✓ |
| AP-CVD (this work) | ✓ | ✓ | ✓ | ✓ | ✓ |

[a] '–' signifies partial fulfilment
[b] Processing window captures the range of deposition temperatures compatible with growing on perovskites, range of oxidants that can be used, extent to which the deposition method damages the perovskite and ultimately the efficiencies that can be achieved

An unexplored alternative for oxide film growth over lead-halide perovskites is atmospheric pressure chemical vapor deposition (AP-CVD). AP-CVD uses the same precursors as ALD, but mixes them in the gas phase to obtain CVD rather than self-limiting growth (**Figure 1**a). Thus AP-CVD has higher growth rates per unit time and per cycle than spatial ALD.[23,26] Also, AP-CVD processes oxides under ambient conditions without a vacuum chamber and is compatible with high-throughput roll-to-roll manufacturing, unlike pulsed-CVD.[23,24,27–29] We previously developed a gas manifold (Vertical Cambridge University Close Proximity reactor, or V-CUCP) that distributes the gas precursors uniformly across the manifold and guides the gases vertically down to the substrate held tens of μm below (Figure 1a). ZnO films grown by AP-CVD using the V-CUCP reactor were deposited at similar temperatures as ALD, with comparable thickness uniformity,[23,24,30,31] conformality to high-aspect ratio nanowires,[24,32] and with growth rates of 1.1 nm s$^{-1}$, which was an order of magnitude larger than by ALD. Although AP-CVD has the potential to fulfil the film quality, growth rate, scalability and conformality requirements for manufacturing oxide buffer layers, a key question is whether the shorter deposition times achievable can allow higher deposition temperatures to be used, and whether this can lead to increased device performance. Furthermore, we have previously only demonstrated the growth of ZnO and $Zn_{1-x}Mg_xO$ on $MAPbBr_3$ for light-emitting diodes.[33] However, ZnO is known to degrade $MAPbI_3$ perovskites due to the high isoelectric point,[34,35] and groups have reported that the diethylzinc precursor itself may react with the perovskite.[26] It is therefore important to understand whether AP-CVD can grow alternative oxides that are chemically inert to iodide-based perovskites and overcome the limitations of the current methods for growing oxide overlayers on lead-halide perovskites.

In this work, we aim to understand whether AP-CVD can be used to broaden the processing window of *n*-type oxides grown over lead-halide perovskite solar cells. For the *n*-type oxides, we investigate $TiO_x$ and $SnO_x$, grown by AP-CVD over perovskites for the first time, and compare with ZnO. We investigate the properties of the oxides grown, and compare their deposition on perovskite solar cells based on thermally-sensitive $MAPbI_3$ and the more thermally-stable triple-cation perovskites ($Cs_{0.05}(MA_{0.17}FA_{0.83})_{0.95}Pb(I_{0.84}Br_{0.16})_3$). For all oxide films, we compare the oxide growth at 100 °C with $O_2$ gas oxidant *vs.* 150 °C with $H_2O$ vapor oxidant. In this work, we refer to these as *mild* and *strong* growth conditions respectively. We then investigate the impact of the oxide films formed under these two conditions on the performance of perovskite solar cells. Both the mild and strong growth conditions are typically inaccessible when using ALD to grow oxides over perovskites, either because chemisorption is not possible (mild conditions),[36] or because the perovskite would degrade (strong conditions). Strong growth conditions are also not accessible by solution-processing, thermal evaporation or sputter-deposition for growing oxide buffer layers on perovskites.

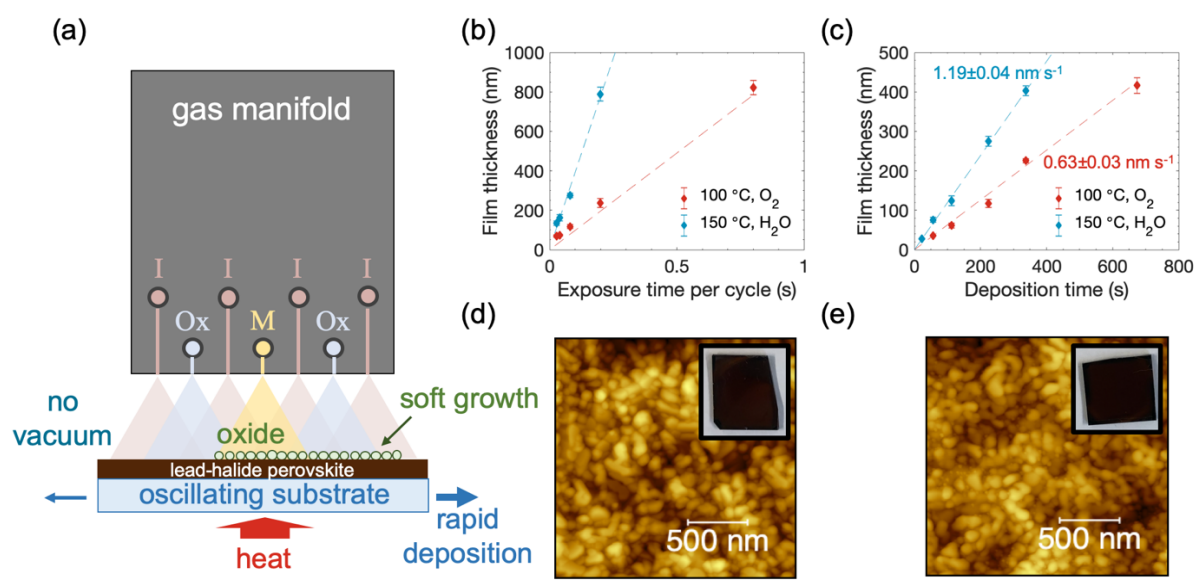

**Figure 1.** (a) Illustration of growing oxides on lead-halide perovskite films by AP-CVD. Metal precursor vapor (M), oxidant (Ox) and inert gas (I, in this case Ar) are introduced to the gas manifold from both sides (cross-section of manifold shown), where they are guided vertically down to the substrate. The metal precursor and oxidant mix in the gas phase, and the oxide thin film is grown by chemical vapor deposition. The oxide thickness depends on the number of times the substrate is moved through the gas channels. The oxide is grown by a soft chemical-based method (unlike sputter-deposition), and the growth rates are high with no vacuum required (unlike ALD and thermal evaporation). Properties of AP-CVD $TiO_x$ grown onto $MAPbI_3$. (b) Thickness of $TiO_x$ against exposure time per AP-CVD cycle, (c) thickness against total deposition time. Morphology, measured by atomic force microscopy, of (d) $MAPbI_3$ thin film deposited on glass and (e) a sister sample with $TiO_x$ grown directly on top at 150 °C using $H_2O$ vapor as the oxidant. The root mean square roughness for both samples was 14 nm. Photographs of the samples measured are inset.

## 2. Results and Discussion

### 2.1. $TiO_x$ Overlayers

We began by focusing on $TiO_x$. By varying the exposure time per cycle in our V-CUCP reactor from 27 ms to 800 ms, we found the growth regime to be CVD rather than ALD based on the linear dependence of the film thickness on exposure time, with no evidence of saturation being reached (Figure 1b; in ALD, the Ti precursor pulse times are typically 100-150 ms).[14,37] Under mild processing conditions, the growth rate was 0.63±0.03 nm s$^{-1}$, while under strong processing conditions, the growth rate was 1.19±0.04 nm s$^{-1}$ (Figure 1c). The growth rates per AP-CVD cycle are shown in Figure S1, SI. The growth rates achieved by AP-CVD for TiO$_x$ are over two orders of magnitude larger than TiO$_x$ grown by ALD at comparable temperatures.[38] Our X-ray photoemission spectroscopy (XPS) measurements of the TiO$_x$ films (Figure S2, SI) had no distinct Cl 2$p$ core level peak, indicating a high reaction yield of the TiCl$_4$ precursor, with insignificant Cl incorporation in the resulting TiO$_x$ films. From XPS measurements of TiO$_x$ grown on gold, we found the $x$ values to be 2.00±0.05 under mild and strong conditions. X-ray diffraction measurements showed the films to be amorphous (Figure S4, SI), which is consistent with ALD films grown at comparable temperatures.[39] The refractive indices of the TiO$_x$ films were 2.2 (mild conditions) and 2.3 (strong conditions) at 640 nm wavelength (Figure S5, SI), which are comparable to the refractive indices of TiO$_2$ grown by ALD under fully-saturated conditions at similar temperatures.[39,40] This indicates that AP-CVD TiO$_x$ was dense and compact, with the films grown under strong conditions being slightly denser. In addition, atomic force microscopy measurements showed the films to be free from pinholes (Figure S6, SI).

To understand whether the growth of AP-CVD TiO$_x$ over perovskites leads to damage, we used strong processing conditions (150 °C, H$_2$O vapor) with the most unstable perovskite (MAPbI$_3$). After we grew 5 AP-CVD cycles of TiO$_x$ (7 nm film thickness, as determined from Figure S1, SI) directly over MAPbI$_3$ thin films, there was no noticeable visual change in the film appearance or morphology, as measured by atomic force microscopy (Figure 1d,e). This suggests that the rapid growth of TiO$_x$ avoided the degradation of thermally-sensitive MAPbI$_3$. This contrasts to previous reports for ALD TiO$_x$, in which a 50 nm thick PC$_{61}$BM layer was required over MAPbI$_3$ to avoid degradation when growing TiO$_x$ on top by ALD, due to the much longer total deposition time of approximately 1 h.[41]

We next focused on the structural and optoelectronic properties of MAPbI$_3$ films after the growth of the AP-CVD TiO$_x$ overlayer. To keep the system investigated comparable to the structure used in photovoltaic devices, we covered the MAPbI$_3$ film with a 5 nm layer of PC$_{61}$BM (thickness measured by atomic force microscopy). In this layout, the samples mimic the layer stack in the solar cells with comparable chemical reactivity between the individual layers. At the same time, the PC$_{61}$BM layer is not sufficiently thick to protect the perovskite from thermal damage, and there would thus be more degradation than in full devices. However, when we measured the diffraction pattern of the PC$_{61}$BM-coated MAPbI$_3$ samples after depositing 5 AP-CVD cycles of TiO$_x$ under mild and strong conditions, the films remained phase-pure with no PbI$_2$ peaks appearing (Figure 2a; close-up in Figure S7, SI). This suggests that the rapid growth of TiO$_x$ by AP-CVD led to an avoidance of bulk structural damage. By contrast the growth of TiO$_2$ with comparable thickness by ALD at only 100 °C has been reported to lead to degradation.[41] We also observed no significant quenching of the photoluminescence (PL) lifetime when TiO$_x$ was grown by AP-CVD over MAPbI$_3$/PC$_{61}$BM under strong processing conditions (Figure 2b). Similarly, TiO$_x$ grown under strong conditions directly over MAPbI$_3$ films did not lead to faster PL decay (Figure S8, SI). These suggest that no extra non-radiative recombination pathways were introduced by AP-CVD growth.

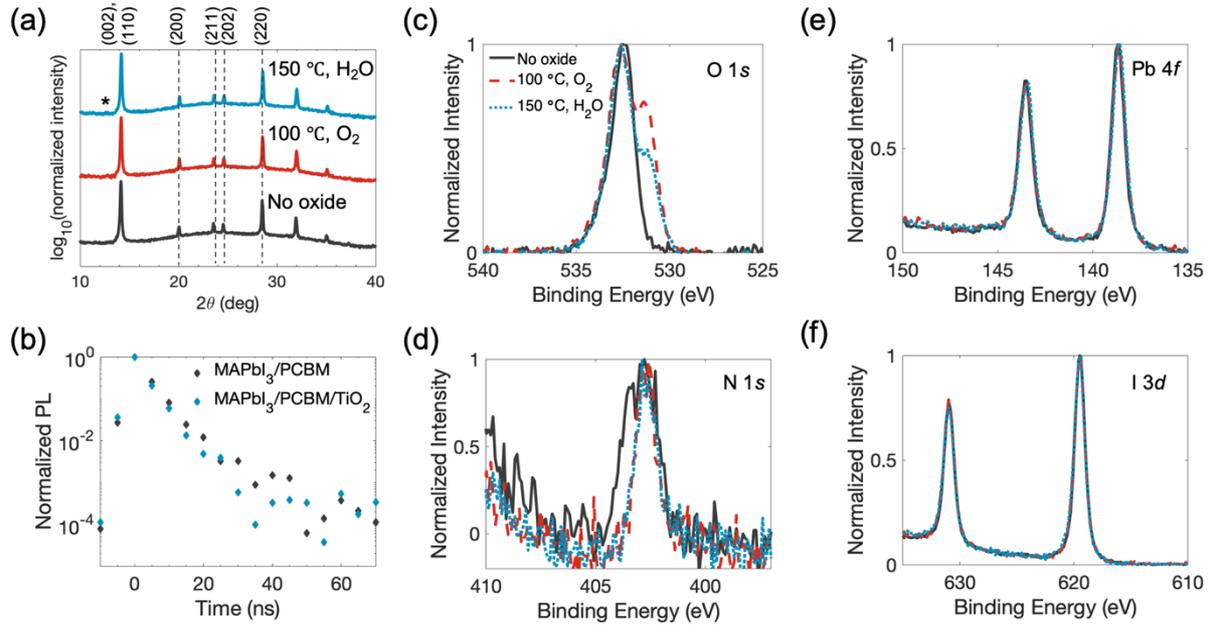

**Figure 2.** Analysis of MAPbI$_3$ films coated with a non-compact 5 nm layer PC$_{61}$BM. (a) Diffraction patterns and (b) time-resolved photoluminescence (PL) of MAPbI$_3$/PC$_{61}$BM with TiO$_x$ grown on top. * denotes the position of where the PbI$_2$ (001) peak would be. X-ray photoemission spectra of (c) O 1$s$, (d) N 1$s$, (e) Pb 4$f$ and (f) I 3$d$ core level regions for MAPbI$_3$/PC$_{61}$BM thin films compared to corresponding samples with TiO$_x$ (5 AP-CVD cycles) grown on top. The legend for parts c) – f) are the same.

To determine whether the growth of the AP-CVD TiO$_x$ overlayer damaged the perovskite surface and to access potential chemical reactions at the interface,[42] we performed X-ray photoemission spectroscopy measurements ($h\nu$ = 1486.6 eV) on a set of sister MAPbI$_3$ samples covered with a 5 nm layer of PC$_{61}$BM. The thin PC$_{61}$BM layer was non-compact and we were able to capture the signal from the perovskite surface. The survey spectra are shown in Figure S3a, SI. Figure 2c shows that the samples with no oxide have an O 1$s$ core peak centered at 532.6 eV (Figure S3b&c, SI), which is attributed to adventitious oxygen and oxygen in the carboxylic acid group of the PC$_{61}$BM molecule.[43] Following the deposition of the TiO$_x$ overlayers, another O 1$s$ core peak centered at a binding energy of 531.1 eV appeared (Figure 2c & Figure S3b&c, SI). This has been attributed to oxygen bonded to titanium.[44] These results are consistent with the Ti 2$p_{3/2}$ core level spectra, which only have peaks for the samples covered with TiO$_x$ (Figure S3d–f, SI). Figure 2d shows the N 1$s$ core levels. The peak centered at 402.5 eV is consistent with nitrogen present in methylammonium.[45] Following TiO$_x$ deposition, no N 1*s* core peaks associated with degradation products from MA$^+$ (which would be located at lower binding energies)[45] were found. The N 1*s* full width at half maximum (FWHM) also did not become broader and the N/Pb ratio did not decrease (Table 2). Both suggest that the thermally-sensitive MA$^+$ cation was not damaged and may not have been significantly removed from the lattice. The inorganic lattice of MAPbI$_3$ also exhibited no damage, since no metallic Pb$^0$ core peaks (Figure 2e) or oxyiodide (Figure 2f) peaks were found. Whilst the I/Pb ratio was lower after the deposition of TiO$_x$ (Table 2), the ratio remains within the 'defect-tolerant' region for MAPbI$_3$.[45] Our XPS results therefore indicate that the surface of MAPbI$_3$ was not thermally damaged after the deposition of AP-CVD TiO$_x$ on the MAPbI$_3$/(5 nm) PC$_{61}$BM stack owing to the rapid growth rates achievable.

**Table 2.** Ratio of elements in MAPbI$_3$ films measured by X-ray photoemission spectroscopy

| Ratio | No Oxide | With TiO$_x$ | |
|---|---|---|---|
| | | 100 °C, O$_2$ gas | 150 °C, H$_2$O vapor |
| N/Pb | 1.1±0.1 | 1.3±0.1 | 1.1±0.1 |
| I/Pb | 3.1±0.1 | 2.7±0.1 | 2.7±0.1 |

To determine the effect of the TiO$_x$ growth conditions on the performance of MAPbI$_3$ solar cells, we fabricated devices using a *p-i-n* structure (Figure 3a), with solution-processed NiO$_x$ as the hole transport layer and 40 nm thick PC$_{61}$BM as the electron transport layer (as detailed in the Experimental Procedures). We used bathocuproine (BCP) and Ag for the top electrode, in which the BCP acted as an interface modifier to ensure no barrier to electron extraction. Devices with a TiO$_x$ overlayer grown under mild conditions not only showed no decrease in performance compared to the control, but rather gave an improvement in efficiency (Figure 3b). Similarly, devices with TiO$_x$ grown under strong conditions gave more efficient devices than the control. We then kept the oxidant used to H$_2$O vapor and varied the TiO$_x$ growth temperature from 60 °C to 200 °C. Surprisingly, we found that there was no significant decrease in the power conversion efficiency compared to the control until the growth temperature was 180 °C (Figure 3b). This is significantly higher than the perovskite-compatible growth temperatures reported for solution processing, ALD or $TiO_2$, in which the oxide layers could only be processed at up to 100 °C (Table SI, SI).[35] We found that the range of perovskite-compatible growth temperatures was larger for triple-cation perovskites. With these samples, the power conversion efficiencies did not significantly decrease until the growth temperature was above 180 °C (Figure 3c). Even for the triple-cation perovskite devices with $TiO_x$ grown at 220 °C, there were some devices that were more efficient than the control (Figure S10, SI). This suggests that despite such elevated processing temperatures, not all of the perovskite device was degraded, which contrasts with perovskite devices that have solution-processed, sputter-deposited or ALD overlayers.[15,41]

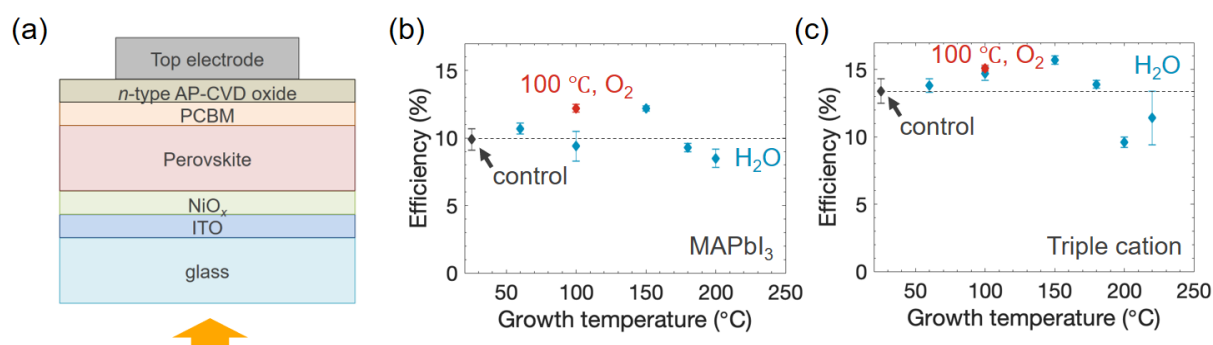

**Figure 3.** (a) Device structure used. The AP-CVD oxide here was $TiO_x$. Power conversion efficiency of (b) $MAPbI_3$ and (c) triple-cation perovskite solar cells as a function of the temperature of the platen the samples were mounted on when depositing an overlayer of $TiO_x$ by AP-CVD. The triple-cation perovskite was $Cs_{0.05}(MA_{0.17}FA_{0.83})_{0.95}Pb(I_{0.84}Br_{0.16})_3$. The blue points are for cells with $TiO_x$ grown under strong conditions. The red point is for cells with $TiO_x$ grown under mild conditions. These are compared to cells with no oxide overlayer (control). In each case, the performance of six samples was averaged. The error bars each represent two standard errors.

Our most efficient devices were triple-cation perovskites that had $TiO_x$ grown at 150 °C using $H_2O$ vapor as the oxidant (Figure 3b&c). When we replaced the BCP/Ag top electrodes with more reflective Al electrodes to obtain a more efficient control device (15.9% power conversion efficiency), adding a $TiO_x$ layer grown under strong conditions improved the performance to 17.1% (details in Figure S11 and Table S2, SI). The improvement in performance over the control was primarily due to an increase in the shunt resistance from $3000 \pm 1000$ $\Omega$ $cm^2$ (control) to 4000±3000 Ω cm² (with TiO$_x$), which correlated with increases in the fill factor from 70±3% to 76.4±0.3%. This is likely due to the spin-coated PC$_{61}$BM layer being non-compact (Figure S12, SI) and the dense AP-CVD TiO$_x$ forming a pinhole-free layer covering it, thereby reducing shunt pathways.

But growing TiO$_x$ at 150 °C also led to more efficient performance compared to other growth temperatures (Figure 3b&c). We hypothesized that the improvement in device performance when increasing the TiO$_x$ growth temperature to 150 °C was due to a reduction in the resistivity of the oxide films. To test this, we grew 300 nm thick AP-CVD TiO$_x$ on ITO and evaporated Al on top. We measured the current density against applied bias and fitted the linear regime to calculate the resistivity (refer to Figure S13, SI for representative measurements). From these measurements, we found that the resistivity decreased from (4.7±0.8) × 10$^4$ Ω cm (100 °C) to (1.7±0.2) × 10$^3$ Ω cm (150 °C), as shown in Figure 4. This correlated with a decrease in the series resistance of the MAPbI$_3$ solar cells from 3.9±0.6 Ω cm² (100 °C) to 2.2±0.2 Ω cm² (150 °C), and an increase in the fill factor from 56±1% (100 °C) to 74±1% (150 °C). Similar trends were observed with the triple-cation perovskite devices. This shows the importance of being able to open-up the processing window for TiO$_x$ in order to achieve higher-performing devices.

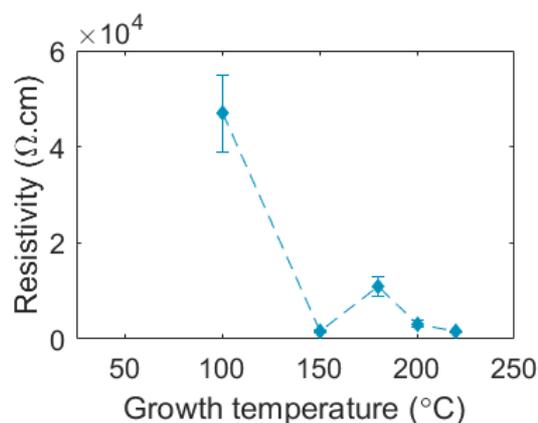

**Figure 4.** Resistivity of TiO$_x$ grown at different temperatures using H$_2$O vapor as the oxidant. We note that the TiO$_x$ films grown at temperatures above 150 °C had comparable or lower resistivity (Figure 4). However, further increases in efficiency were not observed. By contrast

devices with TiO$_x$ grown at 180 °C and higher had larger series resistance and lower fill factors. This is likely due to the thermally-induced degradation of the perovskite, which manifested as an overall reduction in the power conversion efficiency compared to the control when growth temperatures were higher than 180 °C.

We also note that TiO$_x$ films grown at 100 °C using O$_2$ gas as the oxidant have similar resistivity ((4±3) × 10$^4$ Ω cm) as the films grown using H$_2$O vapor ((4.7±0.8) × 10$^4$ Ω cm). However, in both MAPbI$_3$ and triple-cation perovskite devices, the series resistance of the devices with TiO$_x$ grown with O$_2$ gas was lower than those with H$_2$O vapor. This was likely because the TiO$_x$ films were approximately half the thickness, due to the lower growth rate (Figure 1b), since the same number of AP-CVD cycles were used.

## 2.2. ZnO and SnO$_x$ Overlayers

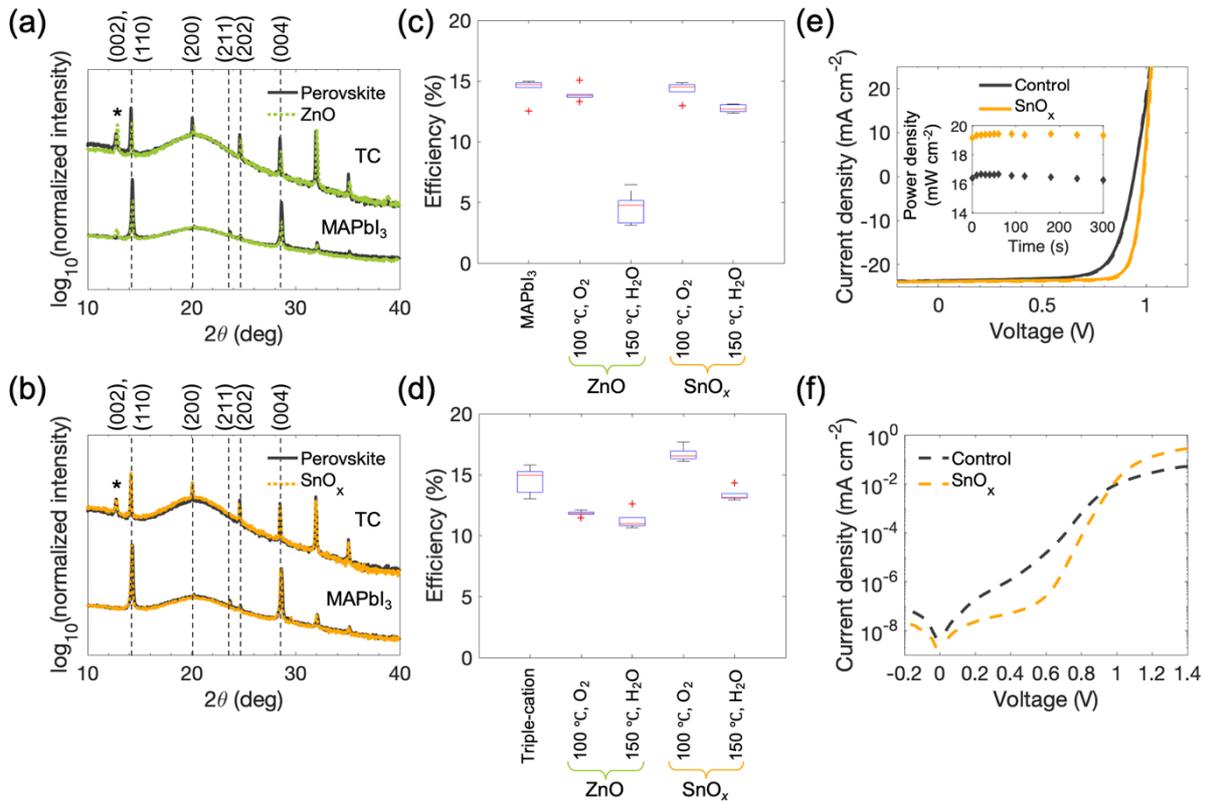

**Figure 5.** Growth of AP-CVD ZnO and SnO$_x$ on MAPbI$_3$ and triple-cation (TC) perovskite. The TC perovskite was Cs$_{0.05}$(MA$_{0.17}$FA$_{0.83}$)$_{0.95}$Pb(I$_{0.84}$Br$_{0.16}$)$_3$. X-ray diffraction patterns of as-deposited perovskite films compared to films that have (a) ZnO or (b) SnO$_x$ overlayers grown at 100 °C with O$_2$ gas as the oxidant. Performance of (c) MAPbI$_3$ or (d) TC solar cells with oxide overlayers. The device structure was

glass/ITO/NiO$_x$/perovskite/(40 nm) PC$_{61}$BM/oxide/Al. (e) Illuminated and (f) dark current-density *vs.* voltage curves and (inset) steady-state power output of triple-cation PVs with AP-CVD SnO$_x$.

To determine whether AP-CVD can be more generally applied to enable other *n*-type oxides to be grown over perovskites across a wider processing window, we investigated ZnO and SnO$_x$. We have previously shown that ZnO deposited with our V-CUCP reactor is grown under CVD mode with a growth rate of 1.1 nm s$^{-1}$.[30] We investigated the growth regime of SnO$_x$ and found it to also be CVD from the linear relationship between film thickness and exposure time (Figure S14a, SI). The growth rates were 0.41±0.04 nm s$^{-1}$ (mild growth conditions) and 0.27±0.02 nm s$^{-1}$ (strong growth conditions), as shown in Figure S14b, SI. These growth rates were two orders of magnitude larger than literature reports for ALD SnO$_2$ grown at 80 °C[25] or 100 °C,[14] as well as being an order of magnitude higher than pulsed-CVD.[14] The films grown were amorphous (Figure S15, SI), and the refractive indices were 1.8 at a wavelength of 632 nm under both growth conditions (Figure S16, SI), which is similar to previous reports of dense ALD SnO$_2$ films grown at these temperatures.[22] Atomic force microscopy measurements of the films showed them to be free from pinholes (Figure S17, SI). The AP-CVD SnO$_x$ films were uniform over the deposition area, with the thickness varying by only 1% (two standard errors compared to the average thickness), as shown in Figure S18, SI. From X-ray photoemission spectroscopy measurements, we found the *x* value for SnO$_x$ grown under both mild and strong conditions to be 1.98±0.07 (Figure S19&S20, SI). We previously found that AP-CVD ZnO grown in the same reactor has a thickness variation of only 3% over the entire deposition area.[41]

We grew 100 AP-CVD cycles of ZnO and SnO$_x$ (2 min total deposition time) directly onto MAPbI$_3$ and triple-cation perovskite films. The control triple-cation perovskite films without oxide overlayers have a small quantity of residual PbI$_2$, which has been commonly observed in previous works.[42,43] The growth of the ZnO overlayers under mild conditions resulted in a higher relative intensity of the PbI$_2$ peak for both MAPbI$_3$ and triple-cation perovskite films (Figure 5a, Figure S21a&b, SI and Figure S22a&b, SI). This may have been due to ZnO itself degrading the perovskite due to its high isoelectric point, as discussed in previous reports.[44,45] But with the SnO$_x$ overlayers, the MAPbI$_3$ films did not form a PbI$_2$ diffraction peak, even under strong growth conditions (Figure 5b & Figure S21c, SI and Figure S22c, SI). For the triple-cation perovskites, the residual PbI$_2$ peak did not increase when SnO$_x$ was grown on top under mild conditions (Figure 5b). However, under strong growth conditions, there was a small increase in the PbI$_2$ peak and small decrease in the intensity of the (002) and (110) perovskite peaks, as well as the appearance of a small peak at 11.7° 2$\theta$ (Figure S21d, SI and Figure S22d, SI), which we attribute to the appearance of the cesium lead halide delta phase.[46] It has previously been reported that at temperatures above 110 °C, the tetrakis(dimethylamido)tin precursor can remove the FA$^+$ cation from the perovskite.[15] This would agree with our observations, in which the deposition of the SnO$_x$ overlayer at 150 °C could have removed a small quantity of FA$^+$, resulting in a small increase in PbI$_2$ content and the formation of a cesium lead halide delta phase due to a more Cs-rich perovskite being present. However, we note that the intensity of both the PbI$_2$ and delta-phase peaks remain small, whereas previous reports of ALD or pulsed-CVD SnO$_2$ grown at 150 °C on Cs-FA perovskites resulted in the PbI$_2$ peak being larger than the perovskite (002) & (110) peaks. We attribute the limited bulk damage to the perovskite to the deposition time being significantly shorter (2 min here *vs.* <15 min for pulsed-CVD and ~60 min for ALD).[15] In addition, our results suggest that MA$^+$ was not as significantly affected by the tin precursor, hence the absence of PbI$_2$ formation in MAPbI$_3$. Overall, our results show the growth of a SnO$_x$ overlayer to not be as damaging to the bare perovskite films as ZnO, thus showing the importance of looking beyond ZnO for AP-CVD oxide overlayers for lead-halide perovskites.

Consistent with our X-ray diffraction measurements, we found that the growth of a ZnO overlayer led to a significant decrease in device performance, especially for MAPbI$_3$ devices with ZnO grown under strong conditions (Figure 5c&d). This is consistent with an increased PbI$_2$ content in the perovskite films following the deposition of the oxide overlayer. By contrast, the growth of the AP-CVD SnO$_x$ overlayer under both mild and strong conditions did not lead to an as significant decrease in performance. This is consistent with the X-ray diffraction measurements showing that only a small increase in PbI$_2$ content occurred following the growth of the SnO$_x$ overlayer on the triple-cation perovskites under strong conditions. Indeed, the growth of SnO$_x$ using mild conditions led to comparable (MAPbI$_3$) or improved (triple-cation perovskite) performance compared to the control devices (Figure 5c&d).

After optimizing the thickness of the SnO$_x$ layer (grown under mild conditions) to 60 nm, we achieved a triple-cation perovskite device with a power conversion efficiency of 19.7% (reverse sweep)/19.3% (forward sweep), which was close to the steady-state efficiency of 19.4% (Figure 5e, inset). This was higher than the performance of the control device (16.5% reverse sweep; 16.3% forward sweep; 16.5% steady-state efficiency). The improvement in efficiency was statistically significant, with the SnO$_x$-coated devices having an average efficiency of 18±1%, compared to 15±1 % for the control. The means were calculated from six devices for each condition, and the average forward and reverse sweep efficiencies were the same. The improvement in performance was primarily due to an increase in the fill factor from 72±2% (control) to 82±2% (60 nm SnO$_x$). This correlated with an increase in the shunt resistance from 1300±600 Ω cm$^2$ (control) to 3000±1000 Ω cm$^2$ (60 nm SnO$_x$). Our dark current-density *vs*. voltage measurements (Figure 5f) are consistent with a reduction in shunting with the SnO$_x$ overlayer, in which the coated triple-cation perovskite devices showed lower reverse-bias dark current density and a higher rectification ratio. This is similar to our results with the TiO$_x$ overlayer and is likely due to the dense oxide covering shunt pathways in the PC$_{61}$BM layer (Figure S12, SI). We note that we only measured the dark current density curves to -0.2 V because these devices exhibited voltage breakdown at more negative biases. The fill factor of our champion device with AP-CVD SnO$_x$ was 84%. By contrast, perovskite devices with oxide buffer layers grown by solution-processing and sputter deposition have lower fill factors because of current leakage through less dense films (solution-processing),[48] or because of damage to the perovskite surface and more resistive oxides grown under sub-optimal conditions (sputter deposition).[17] To our knowledge, the champion steady-state efficiency of 19.4% is currently the highest reported for a perovskite device with an oxide buffer layer deposited by ALD, solution processing or sputter deposition (Table S1, SI).

## 2.3. Broad Applications of AP-CVD Oxides in Devices

To investigate whether we can apply our AP-CVD SnO$_x$ to cover textured perovskite films, we grew wrinkled Cs$_{0.17}$FA$_{0.83}$Pb(I$_{0.83}$Br$_{0.17}$)$_3$ perovskites (Figure 6a). These thin films wrinkle during annealing due to compressive stresses,[47] which is achievable by tuning the fraction of dimethyl sulfoxide in the perovskite solvent. After depositing PC$_{61}$BM by solution processing, clusters and streaks formed (Figure 6b), suggesting non-uniform coverage of the PC$_{61}$BM. Depositing 20 nm SnO$_x$ on top resulted in the same morphology as the PC$_{61}$BM-coated sample, suggesting conformal coverage (Figure 6c). We imaged a cross-section of the sample with PC$_{61}$BM and SnO$_x$, and found that while the PC$_{61}$BM did not cover some regions of the wrinkled perovskite, the SnO$_x$ layer was conformal over the entire surface of the sample (Figure S23, SI). The devices with only a PC$_{61}$BM layer exhibited significant shunting (Figure 6c,d), with fill factors of 38±2% and shunt resistances of only 300±50 Ω cm$^2$. However, after the growth the SnO$_x$ overlayer, the fill factor increased and reverse-bias dark current-density decreased with increasing thickness (Figure 6c,d). Through the use of AP-CVD SnO$_x$, we were able to increase the power conversion efficiency from 6.5±0.6% (no oxide; forward sweep) to 10.9±0.5% (115 nm SnO$_x$; forward sweep). This has potential future applications in conformally covering evaporated perovskite films grown over textured silicon cells for tandem photovoltaic devices, which are becoming increasingly popular because they have led to the most efficient perovskite-silicon tandems, and industry-standard silicon solar cells are front-textured to reduce light scattering.[48,49] We note that the AP-CVD $SnO_x$ and $TiO_x$ films exhibit high transmittances >90% in the visible light wavelength range, rising to >98% in the near-infrared wavelength range (Figure S24a&b, SI). This makes them suitable as buffer layers for a semi-transparent perovskite top-cell that have minimal contribution to parasitic optical losses. Finally, we note that when we grew 60 nm $SnO_x$ over (Cs,FA)-based perovskites with a more optimized composition and device structure, with less wrinkling, we improved the median efficiency from 14.7% (no $SnO_x$) to 15.4% (with $SnO_x$; peak value of 16.8%), as detailed in Figure S25, SI.

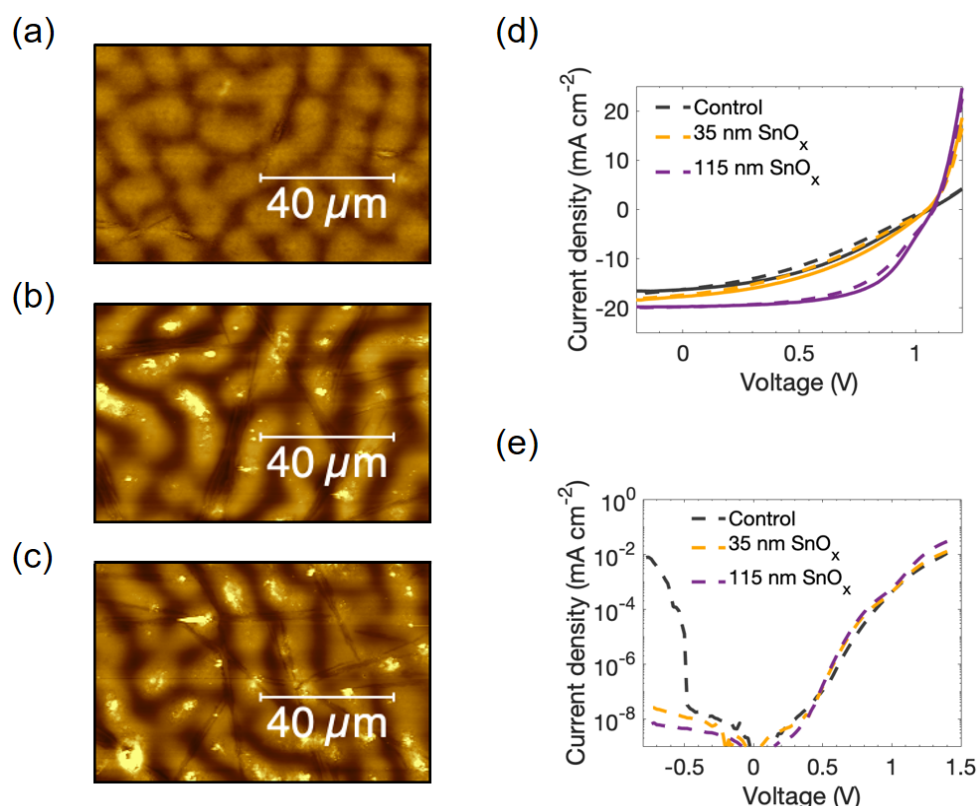

**Figure 6.** Atomic force microscopy image of wrinkled $Cs_{0.17}FA_{0.83}Pb(I_{0.83}Br_{0.17})_3$ thin film (a) with no overlayer, (b) coated with $PC_{61}BM$ and (c) coated with $PC_{61}BM$ and $SnO_x$. The color scales in all atomic force microscopy images were adjusted to range from -400 nm to 400 nm. (d) Illuminated current density–voltage curves and (e) dark current density–voltage curves of $Cs_{0.17}FA_{0.83}Pb(I_{0.83}Br_{0.17})_3$ devices coated with $PC_{61}BM$ and with no oxide (control), 35 nm and 115 nm $SnO_x$.

## 3. Conclusion

Growth of oxide buffer layers over lead-halide perovskites and their organic charge transport layers is critically important for protecting against environmental and mechanical damage, and requires finely-tuned processing parameters compatible to the thermally-unstable perovskite layer. In this work, we have shown AP-CVD to overcome the limitations of the current methods to fulfil all key requirements for growing oxide buffer layers on lead-halide perovskites. AP-CVD films are pinhole-free and uniform over large area, can be grown rapidly without vacuum, and are conformal to textured perovskites. Using $TiO_x$ as the main test material ($x = 2.00\pm0.05$), we found that the short processing times allowed us to broaden the range of deposition temperatures compatible with the perovskite stack from <110 °C (spin-coating, ALD, sputter-deposition and thermal evaporation) to ≥180 °C (AP-CVD). $TiO_x$ grown at higher temperature has lower resistivity, resulting in efficiencies improving by >2% in single-junction perovskite solar cells.

More broadly, we showed that AP-CVD can also grow uniform $SnO_x$ ($x = 1.98\pm0.07$) and ZnO on perovskites, but the ZnO degraded the perovskite, emphasizing the importance of looking beyond ZnO for buffer layers. Using pinhole-free AP-CVD $SnO_x$ to cover the non-continuous $PC_{61}BM$ layer in triple-cation perovskite devices, we reduced current leakage, resulting in fill factors reaching 84% and steady-state efficiencies increasing from 16.5% (no $SnO_x$) to 19.4% (with 60 nm $SnO_x$). To our knowledge, 19.4% is currently the highest power conversion efficiency amongst *p-i-n* structured perovskite solar cells with oxide buffer layers grown by any practical coating method. AP-CVD is therefore a versatile technique that is appealing for manufacturing oxide buffer layers for perovskite optoelectronic devices.

**Author Contributions**

R.D.R. and R.A.J. contributed equally to this work. R.L.Z.H. conceived of the project, fabricated the perovskite films and devices, performed the profilometry measurements on the oxide films, performed the X-ray diffraction and time-resolved photoluminescence measurements. R.D.R. developed the growth of the oxide films by AP-CVD. R.A.J. grew the oxide films and performed the spectroscopic ellipsometry measurements with the help of T.N.H. S.B., M.B., M.F. and P.S. performed and analyzed the photoemission spectroscopy measurements. T.N.H. performed the atomic force microscopy measurements. M.N. performed the current-voltage measurements for determining the oxide film resistivity. Y.-H.L., M.S. and W.L. contributed to fabricating the perovskite devices and analysis. Z.L. performed the UV-visible spectrophotometry measurements. All authors contributed to writing the paper and discussing the results.

**Acknowledgements**

The authors would like to thank Prof. Richard Friend for useful discussions on this project. The authors also thank Shahab Akhavan for technical assistance in evaporating Al electrodes onto the oxides for resistivity measurements and Yu-Hsien Chiang for discussions on perovskite fabrication. R.D.R. and R.A.J. acknowledge support from DTP studentships funded by the EPSRC (No.: EP/N509620/1). R.A.J. and J.L.M.-D. thank Bill Welland for financial support. R.L.Z.H. acknowledges support from the Royal Academy of Engineering under the Research Fellowship scheme (No.: RF\201718\1701), the Isaac Newton Trust (Minute 19.07(d)), the Centre of Advanced Materials for Integrated Energy Systems (CAM-IES; EP/P007767/1), and Magdalene College Cambridge. We also acknowledge support from the EPSRC (Nos.: EP/P027032/1 and EP/M005143/1), as well as from the Winton Programme for the Physics of Sustainability under the Pump-Prime scheme. This work was partly carried out in the framework of a project of IPVF, which was supported by the French Government as part of their programme of investment in the future (Programme d'Investissement d'Avenir ANR-IEED-002-01). P.S. thanks the French Agence Nationale de la Recherche for funding under the contract number ANR-17-MPGA-0012. T.N.H. was supported by the EPSRC Centre for Doctoral Training in Graphene Technology (No.: EP/L016087/1). M.N. acknowledges financial support from the European Commission through a Marie-Curie Individual Fellowship (EC Grant Agreement Number: 747461). W.L. acknowledges the National Natural Science Foundation of China (no. 61805203).